\documentclass[aps,english,prb,floatfix,superscriptaddress,twocolumn,amsmath,amssymb,longbibliography, reprint]{revtex4-2}

\usepackage{xcolor}
\usepackage{textcomp}
\usepackage{amsmath}
\usepackage{amssymb}
\usepackage{graphicx}
\usepackage{wasysym}
\usepackage{gensymb}

\makeatletter

\providecommand{\tabularnewline}{\\}

\usepackage{dcolumn}
\usepackage{bm}
\usepackage{braket}
\usepackage{hyperref}
\usepackage{booktabs}
\usepackage[dvipsnames]{xcolor}
\usepackage{hyperref}
\hypersetup{
    breaklinks=true,
    colorlinks=true,
    linkcolor=blue,
    urlcolor=blue,
    citecolor=blue,
    filecolor=blue,
    pdftitle={},
    pdfauthor={}
}

\usepackage{bbold}
\usepackage{dsfont}
\PassOptionsToPackage{caption=false}{subfig} 

\IfFileExists{lmodern.sty}{\usepackage{lmodern}}{}

\definecolor{orange}{RGB}{255,127,0}

\usepackage{braket}

\makeatother

\begin{document}

\title{Modeling strained Cd$_3$As$_2$ thin films and their behavior in magnetic fields}

\author{M. Smith}
\affiliation{Materials Science Division, Argonne National Laboratory, Lemont, Illinois 60439, USA}
\author{A.A. Burkov}
\affiliation{Department of Physics and Astronomy, University of Waterloo, Waterloo, Ontario N2L 3G1, Canada}
\affiliation{Perimeter Institute for Theoretical Physics, Waterloo, Ontario N2L 2Y5, Canada}
\author{P.~P. Orth}
\affiliation{Ames National Laboratory, Ames, Iowa 50011, USA}
\affiliation{Department of Physics, Saarland University, 66123 Saarbr\"ucken, Germany}
\affiliation{Center for Quantum Technologies (QuTe), Saarland University,
66123 Saarbr\"ucken, Germany}
\author{I. Martin}
\affiliation{Materials Science Division, Argonne National Laboratory, Lemont, Illinois 60439, USA}
\author{Victor L. Quito}
\email{vquito@ifsc.usp.br}
\affiliation{S\~{a}o Carlos Institute of Physics, University of S\~{a}o Paulo,
PO Box 369, 13560-970, S\~{a}o Carlos, SP, Brazil}
\affiliation{Ames National Laboratory, Ames, Iowa 50011, USA}

\date{\today}

\begin{abstract}
We present a systematic analysis of the behavior of thin films of Cd$_3$As$_2$ under different strain profiles and in magnetic fields. In each case, we construct effective $k \cdot p$ models by considering the reduction of symmetry and all constraints imposed by the remaining symmetries. Our analysis naturally describes both in-plane biaxial and uniaxial strain. Biaxial strain is expected to preserve in-plane $C_4$ rotational symmetry while breaking inversion, allowing for a description in terms of the $4mm$ point group. Uniaxial strain, on the other hand, breaks $C_4$ symmetry. For this case, we consider two scenarios: one preserving inversion, described by the $mmm$ group, and one breaking it, leading to $2mm$ symmetry. After deriving the models, we examine the effects of out-of-plane magnetic fields, identifying two possible microscopic mechanisms that can account for the experimental results reported in Ref.~\cite{ahadi_strain-induced_2025}. Importantly, our analysis proposes a new method for differentiating between them. By incorporating the effects of multiple subbands along the confinement direction, we show that the opening of a gap in the lowest Landau level requires either reducing the symmetry down to $2mm$, breaking both inversion and $C_4$ rotations, or a topological transition of the band structure due to strain-induced band renormalization. Furthermore, we demonstrate that a two-dimensional Dirac semimetal phase can be induced by sufficiently large in-plane magnetic fields. This phase is highly sensitive to different strain profiles, with band touchings occurring when the field is applied perpendicular to preserved mirror planes, serving as a powerful probe of the material's strain profile.
   
\end{abstract}
\maketitle

\addtolength{\abovedisplayskip}{-1mm}

\section{Introduction~\label{sec:Intro}} 

Cd$_3$As$_2$ has recently garnered significant experimental and theoretical attention for being a highly tunable and clean experimental realization of different semimetallic and insulating topological phases and its promising applications as THz radiation source and in high-frequency transistors~\cite{crassee_3d_2018,shoron_prospects_2020,cheng_efficient_2020,yang_broadband_2022}. 
In the bulk, it is a three-dimensional topological Dirac semimetal ~\cite{wehling_dirac_2014,wang_three-dimensional_2013,yang_classification_2014,neupane_observation_2014,liu_stable_2014,yi_evidence_2014,borisenko_experimental_2014,jeon_landau_2014,he_quantum_2014,zhang_breakdown_2015,roth_reinvestigating_2018} with nodal points that are protected by fourfold rotation symmetry. In thin film geometries, it is a realization of a topological quantum spin Hall (QSH) insulator for samples oriented along the (001) direction~\cite{guo_hall_2022,Stemmer_PRLa_2023,Stemmer_PRLb_2023}. Furthermore, its response to in- and out-of-plane magnetic fields is intriguing: experiments have shown that it can be tuned to a semimetallic phase upon the application of strong enough in-plane magnetic fields~\cite{Stemmer_PRLa_2023,Stemmer_PRLb_2023}, which has been explained theoretically via an emergent transition to a 2D Dirac phase when a sufficiently strong magnetic field is applied normal to a high symmetry plane~\cite{Stemmer_PRLb_2023,Smith_PRB_2024,miao_engineering_2024}. This makes Cd$_3$As$_2$ a uniquely tunable low-dimensional material with multiple experimentally accessible ways to control the low-energy electronic behavior.

The application of lattice strain is another experimental tuning parameter, which has been theoretically and experimentally studied in bulk Cd$_3$As$_2$ samples~\cite{villar_arribi_topological_2020,pardue_controlling_2021,krizman_enhanced_2022}. While uniaxial strain profiles generally break the fourfold rotation symmetry, it has been experimentally shown that for tensile \emph{biaxial} in-plane strain, the fourfold rotation symmetry can be preserved and the relevant point group is reduced from $4/mmm$ to $4mm$ but remains tetragonal~\cite{pardue_controlling_2021}. 
Thin films under tensile strain can break inversion symmetry, however, as shown in~\cite{ahadi_strain-induced_2025}. This inversion-symmetry breaking was understood in terms of vacancy ordering, which leads to a structure with lower energy, in agreement with DFT calculations.

In the same work, it was also found that under the application of biaxial tensile strain, the crossing of the zeroth Landau levels as a function of magnetic field disappears~\cite{ahadi_strain-induced_2025}. The crossing of the zeroth Landau levels (or lowest Landau levels (LLL)) has been used as an important diagnostic of the nontrivial topology of the relevant bands~\cite{konig_quantum_2008}. However, it is well known that lowering the symmetry group in a QSH insulator can open an anticrossing gap in the zeroth Landau levels, and has been studied in HgTe/CdTe quantum wells~\cite{tarasenko_split_2015,LiuDai_PRB_2025}. The concomitant effects of strain and magnetic fields in generic two-dimensional topological insulators were recently studied, with applications to Cd$_3$As$_2$~\cite{LiuDai_PRB_2025}. There, an analysis based on the orbitals close to the Fermi level and the downfolding of the high-energy bands was performed. Here, we derive effective low-energy models of Cd$_3$As$_2$ in different strain profiles - both biaxial and uniaxial - based on the underlying point group symmetry. We properly take the coupling between different confinement-induced subbands into account and explore the behavior for both in- and out-of-plane magnetic fields. We demonstrate two distinct scenarios that lead to a strain-induced removal of the characteristic crossing of the LLLs as a function of out-of-plane magnetic fields, observed in Ref.~\cite{ahadi_strain-induced_2025}. We also propose an experimental protocol to distinguish the two cases by varying the strain amplitude. For in-plane fields, we discuss how the presence of mirror planes is connected to emergent Dirac nodes at high fields, provided the magnetic field points normal to the mirror plane. We suggest that this behavior can be used to experimentally probe the presence or absence of vertical mirror planes in strained samples.  

\begin{figure}[t]
    \centering
    \includegraphics[width=0.7 \linewidth]{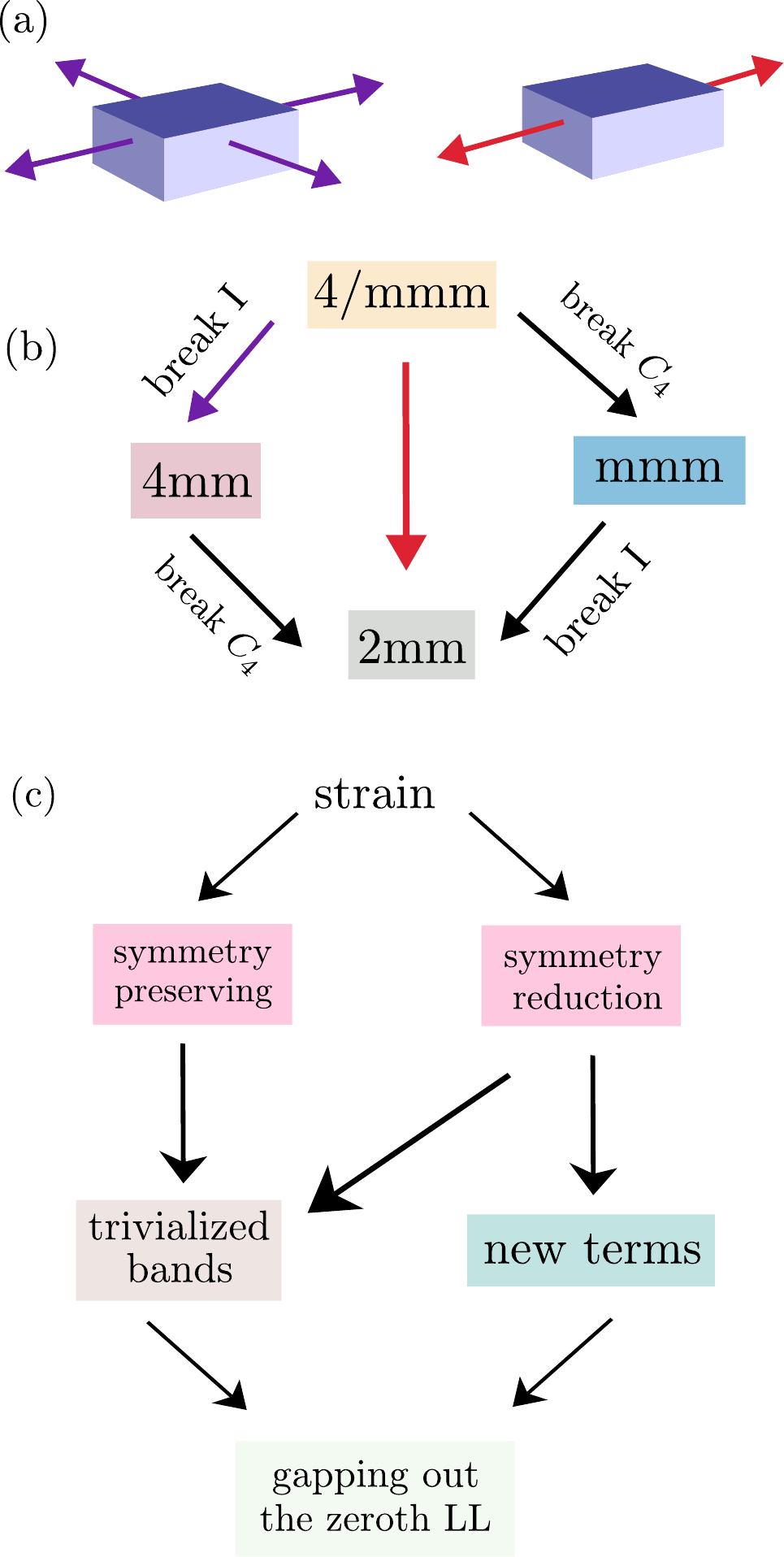}
    \caption{(a) Different strain profiles on a thin film Cd$_3$As$_2$ sample: in-plane biaxial strain (left), which preserves $C_4$ rotations and $C_4$-breaking uniaxial strain (right). (b) Group theory analysis of the effects of the strain on the sample, listing different point groups and the symmetries that are broken. (c) Possible explanations for the experimental results of Ref.~\cite{ahadi_strain-induced_2025}. The first possibility is that strain renormalizes bandstructure parameters and uninverts the bands at the $\Gamma$-point, trivializing the bands. The LLL crossing in magnetic field is then no longer protected and is gapped out. The second possibility is that strain lowers the symmetry by breaking $C_4$ and/or inversion symmetries, leading to new terms that can gap out the LLL crossing. For purely out-of-plane magnetic fields, we demonstrate that the LLL crossing remains unless \emph{both} $C_4$ and inversion are broken. }
    \label{fig:CdAsStrain}
\end{figure}

Our arguments to explain the experimental results of Ref.~\cite{ahadi_strain-induced_2025} are based purely on the existing lattice symmetries, combined with a proper characterization of the subband mixing. Specifically, we perform a systematic analysis for two types of strain in $(001)$ thin-film Cd$_3$As$_2$: biaxial in-plane strain, which preserves the fourfold rotational symmetry $(C_4)$, and uniaxial, $C_4$-breaking strain. The two types of strain are depicted in Fig.~\ref{fig:CdAsStrain}\hyperlink{fig:CdAsStrain}{(a)}. The application of strain leads to a reduction of the point group symmetry from 4/mmm, as shown in Fig.~\ref{fig:CdAsStrain}\hyperlink{fig:CdAsStrain}{(b)}. The application of $C_4$-preserving, but inversion-breaking, biaxial strain leads to a reduction in symmetry from 4/mmm to 4mm. In contrast, uniaxial strain breaks $C_4$ and leads to mmm point symmetry if inversion is preserved or to 2mm if both $C_4$ and inversion are broken. In each case, we investigate the electronic band structure in the presence of in- and out-of-plane magnetic fields and reveal two possible explanations for the LLL anticrossing observed in Ref.~\cite{ahadi_strain-induced_2025}: (1) a strain-induced renormalization of lattice parameters, while keeping the same underlying symmetries, can lead to a topological transition of the electronic bands, or (2) a breaking of $C_{4}$ rotation and inversion symmetries. We show that lifting the crossing of the LLL as a function of magnetic field requires breaking \emph{both} $C_4$ and inversion symmetry, and thus an anticrossing cannot be opened by the application of purely biaxial strain. Considering in-plane magnetic fields, 
we show in Sec.~\ref{sec:inplane_fields} that a semimetallic phase with nodal Dirac points reemerges for large in-plane fields even for broken $C_4$ and inversion. This occurs provided the zero-field bandstructure is topological, and the field is directed perpendicular to the remaining mirror planes. The emergence of a semimetallic phase at large in-plane fields can thus confirm the symmetry reduction scenario. Notably, we find emergent band touchings at large fields whenever the field is directed normal to a mirror plane. The removal of mirror planes by strain can gap the emergent high-field 2D Dirac semimetal phase completely, when the field is applied perpendicular to broken mirror planes, with a gap that depends on the applied strain. We suggest that this prediction can be used as a sensitive probe of existing mirror planes under strain in this material. 

This article is organized as follows. In Sec.~\ref{sec:unstrained}, we review the low-energy effective model of unstrained Cd$_3$As$_2$ and its response to magnetic fields. In Sec.~\ref{sec:biaxial_strain}, we discuss the case of biaxial strain, deriving the additional terms in the effective model that arise from the lowering of the point symmetry. We then investigate the behavior of the strained system in the presence of out-of-plane magnetic fields. In Sec.~\ref{sec:C4_breaking}, we consider strain profiles that break $C_{4}$ symmetry, but preserve inversion, and show that the crossing of the LLL still survives. We then discuss the case of $C_4$ and inversion breaking strain in Sec.~\ref{sec:break_C4_inversion}, where we demonstrate that breaking both symmetries leads to an anticrossing in the LLL spectrum. In Sec.~\ref{sec:inplane_fields}, we address the effects of in-plane fields on different types of strain and reveal a pinning of emergent 2D Dirac nodal points to mirror planes in the system. Lastly, Sec.~\ref{sec:Conclusions} summarizes our results and discusses promising future directions.

\section{Thin film models in magnetic fields~\label{sec:unstrained}}
In this section, we derive effective $k \cdot p$ models for Cd$_3$As$_2$ thin films in magnetic fields, considering both the unstrained case as well as features that occur due to additional terms in cases of biaxial and uniaxial strains. We show how to properly treat the coupling of different confinement-induced subbands arising from $k_z$-dependent terms in the model, and demonstrate that the naive replacement of $k_z \rightarrow n \pi/L$ leads to incorrect results. Finally, we estimate the strain-induced band parameter renormalizations and conclude that strain may be able to trivialize the bands. This provides a possible explanation of the observed LLL anticrossing in biaxially strained samples reported in Ref.~\cite{ahadi_strain-induced_2025} that does not require the breaking of $C_4$ symmetry. 

\subsection{Unstrained thin film model}
We start by reviewing the well-established effective low-energy model for unstrained Cd$_3$As$_2$ thin films. The relevant point group describing the low-energy bands close to the $\Gamma$-point in the Brillouin zone is $\text{4/mmm}$,
and the $k\cdot p$ Hamiltonian for the bulk system, up to quadratic order in $k_x, k_y, k_z$, reads~\cite{Smith_PRB_2024,villar_arribi_topological_2020} 

\begin{equation}
\mathcal{H}^{\text{4/mmm}}=\mathcal{H}^{\left(0\right)}+\mathcal{H}^{\left(1\right)}+\mathcal{H}^{\left(2\right)}, \label{eq:basic_CdAs}
\end{equation}
with
\begin{align}
\mathcal{H}^{\left(0\right)} & =C_{0}\sigma_{0}\tau_{0}+M_{0}\sigma_{0}\tau_{z},\\
\mathcal{H}^{\left(1\right)} & =A\left(k_{x}\sigma_{z}\tau_{x}-k_{y}\sigma_{0}\tau_{y}\right),\\
\mathcal{H}^{\left(2\right)} & =\epsilon_{0}\left(k_x, k_y, k_z\right)\sigma_{0}\tau_{0}+\mathcal{M}(k_x, k_y, k_z)\sigma_{0}\tau_{z}\,.
\end{align}
Here, we have defined the functions 
\begin{align}
\epsilon_{0}\left(k_x, k_y, k_z\right) & =C_{1}k_{z}^{2}+C_{2}\left(k_{x}^{2}+k_{y}^{2}\right),\\
\mathcal{M}(k_x, k_y, k_z) & =-M_{z}k_{z}^{2}-M_{xy}\left(k_{x}^{2}+k_{y}^{2}\right).
\end{align}
The Hamiltonian is written in the $\sigma \otimes \tau$ basis $(\ket{S, 1/2}, \ket{P,3/2}, \ket{S,-1/2}, \ket{P,-3/2})$. In Table~\ref{tab:symm_operat} we list the matrix structure of all the symmetry operations in $\text{4/mmm}$ as well as which symmetries are broken in the relevant subgroups when considering different (biaxial or uniaxial) strain fields. Here and in the following sections, we do not include cubic terms in $\mathbf{k}$ in the models. We have verified that such terms, in general, do not modify our conclusions, and we will comment on their effects when relevant. 

\begin{center}
\begin{table*}
\begin{centering}
\begin{tabular}{|c|c|c|c|c|c|c|}
\hline 
Symmetry & Matrix structure & 4/mmm & 4mm & mmm & 2mm\tabularnewline
\hline 
\hline 
$\ensuremath{C_{2x}}$ & $\ensuremath{-i\sigma_{x}\tau_{z}}$ & $\checkmark$ & $\XBox$ & $\checkmark$ & $\XBox$\tabularnewline
\hline 
$\ensuremath{C_{2y}}=C_{4z}C_{2x}C_{4z}^{-1}$ & $\ensuremath{-i\sigma_{y}\tau_{0}}$ & $\checkmark$ & $\XBox$ & $\checkmark$ & $\XBox$\tabularnewline
\hline 
$\ensuremath{C_{2z}}=C_{4z}^{2}$ & $-i\sigma_{z}\tau_{z}$ & $\checkmark$ & $\checkmark$ & $\checkmark$ & $\checkmark$\tabularnewline
\hline 
$I$ & $\sigma_{0}\tau_{z}$ & $\checkmark$ & $\XBox$ & $\checkmark$ & $\XBox$\tabularnewline
\hline 
$C_{4z}$ & $\frac{1}{\sqrt{2}}\left(\sigma_{0}\tau_{z}-i\sigma_{z}\tau_{0}\right)$ & $\checkmark$ & $\checkmark$ & $\XBox$ & $\XBox$\tabularnewline
\hline 
$T$ & $i\sigma_{y}\tau_{0}K$ & $\checkmark$ & $\checkmark$ & $\checkmark$ & $\checkmark$\tabularnewline
\hline 
$M_{x}$ & $\ensuremath{i\sigma_{x}\tau_{0}}$ & $\checkmark$ & $\checkmark$ & $\checkmark$ & $\checkmark$\tabularnewline
\hline 
$M_{y}=C_{2z}M_{x}$ & $i\sigma_{y}\tau_{z}$ & $\checkmark$ & $\checkmark$ & $\checkmark$ & $\checkmark$\tabularnewline
\hline 
$\ensuremath{M_{+}}=C_{4z}^{-1}M_{x}C_{4z}$ & $-\frac{i}{\sqrt{2}}\left(\sigma_{x}\tau_{z}+\sigma_{y}\tau_{0}\right)$ & $\checkmark$ & $\checkmark$ & $\XBox$ & $\XBox$\tabularnewline
\hline 
$\ensuremath{M_{-}}=C_{4z}M_{x}C_{4z}^{-1}$ & $-\frac{i}{\sqrt{2}}\left(\sigma_{x}\tau_{z}-\sigma_{y}\tau_{0}\right)$ & $\checkmark$ & $\checkmark$ & $\XBox$ & $\XBox$\tabularnewline
\hline 
$\ensuremath{M_{z}}=PC_{2z}$ & $-i\sigma_{z}\tau_{0}$ & $\checkmark$ & $\XBox$ & $\checkmark$ & $\XBox$\tabularnewline
\hline 
\end{tabular}
\par\end{centering}
\caption{The matrix structure of the symmetry operators and the symmetries preserved by each point group relevant to this work. In all cases, time reversal is assumed to be preserved. For time-reversal symmetry, $K$, denotes the complex conjugation.} \label{tab:symm_operat}
\end{table*}
\par\end{center}
In a thin-film geometry with normal along the $z$ direction, we consider the wavefunctions to obey hard-wall boundary conditions at $z=-L/2$ and $z=L/2$,
\begin{align}
\label{eq:subband_Psi_abn}
    \Psi_{\alpha\beta n}(z) &=  \psi_{\alpha\beta}\varphi_{n}(z), 
\end{align}
with $\alpha, \beta = 1,2$ and
\begin{align}
    \varphi_{n}(z) = \sqrt{\frac{2}{L}} \sin\left[\frac{n\pi}{L}\left(z+\frac{L}{2}\right)\right]\,.
    \label{eq:subbands_eigenf}
\end{align}
Here, $n>0$ is a positive integer labeling the confinement-induced subbands, while $\psi_{\alpha\beta}$ is a spinor in pseudospin and orbital space. The effective Hamiltonian for a given subband $n$ as a function of $\mathbf{k} = (k_{x}, k_{y})$ is obtained from Eq.~\eqref{eq:basic_CdAs} by using $\hat{k}_{z} = -i\partial_{z}$, leading to
\begin{align}
    \mathcal{H}^{\text{4/mmm}}_n &= \left(\begin{array}{cccc}
         \mathcal M_n(\mathbf{k}) & A k_+ & 0 & 0 \\
         A k_- & - \mathcal M_n(\mathbf{k}) & 0 & 0\\
         0 & 0 & \mathcal M_n(\mathbf{k}) & - A k_-\\
         0 & 0 & - A k_+ & -\mathcal M_n(\mathbf{k})
    \end{array}\right),
\end{align}
 with $k_{\pm}=k_x\pm i k_y$ and 
 \begin{equation}
     \mathcal M_n(\mathbf{k}) = M_n -M_{xy}(k_x^2 + k_y^2)\,,
 \end{equation}
which contains a subband-dependent mass (or bandgap) 
\begin{align}
    M_n &= M_0 - M_z \left( \frac{n\pi}{L}\right)^2 \label{eq:Mass_n}\,.
\end{align}
Since Eq.~\eqref{eq:basic_CdAs} is diagonal in the subbands $n$ and involves only second derivatives in $z$, there are no terms that mix different subbands up to this order in $(k_x, k_y, k_z)$. We note that even in cubic order we do not find any subband-mixing terms at $k_x = k_y = 0$.

\subsection{Treatment of subband couplings}
While the unstrained model is diagonal in the subband index $n$, we will later see that in strained samples with lower symmetry, additional terms with odd powers in $k_{z}$ appear that couple different subbands. A proper way of treating such terms is by systematic perturbation theory: it is not sufficient to simply replace $k_z$ by $n\pi/L$. We provide a detailed perturbative analysis in Appendix~\ref{sec:App_Subbands} and present the main conclusions here. Obviously, these considerations are not restricted to the case of Cd$_3$As$_2$, but apply to any material in thin film geometry.

Let us illustrate the treatment of the coupling of different subbands with a term of the form 
\begin{equation}
    \mathcal{V}_{1}=A_{z}k_{z}\sigma_{x}\tau_{x}
\end{equation}
with real constant $A_{z}$. Such a term is only allowed when $C_4$ is broken, but is consistent with inversion being preserved, i.e., it appears, for example, in the model for the point group $\text{mmm}$. Given the form of the subband eigenfunctions in Eq.~\eqref{eq:subbands_eigenf}, the operator $\hat{k}_{z}=-i\partial_{z}$ can only connect even to odd (or odd to even) subbands. Thus projecting $\mathcal{V}_1$ onto a fixed subband $n$ in first-order perturbation theory, we find

\begin{equation}
    \left\langle k_{z}\tau_{x}\sigma_{x}\right\rangle_n =\left\langle \tau_{x}\sigma_{x}\right\rangle \int_{0}^{L}dz\varphi_{n}^{*}(z)\left(-i\partial_{z}\right)\varphi_{n}(z)=0\,.
\end{equation}
In general, the expectation value $ \left\langle \tau_{x}\sigma_{x}\right\rangle $ can depend on the subband $n$, as different subbands may have distinct spinor eigenvectors. However, in the 4/mmm point group, which is the basis on which we construct the perturbation theory, the spinor structure is the same for all subbands.  The second equality follows from the orthogonality relation of sines and cosines. Higher-order contributions are non-vanishing, but generally suppressed by energy denominators. For example, in second-order perturbation theory (see Appendix~\ref{sec:App_Subbands} for details), we find a contribution for a generic subband 
\begin{equation}
\label{eq:delta_V1}
    \delta \mathcal{V}_{1} \propto \,-\frac{A_{z}^{2}}{M_{z}}\;\sigma_{0} \tau_{z}.
\end{equation}
The prefactor will depend on the subband index and, in this work, we will focus on the $n=2$ case, as this is the subband closest to the Fermi energy for the sample thickness we are considering, following the experimental values of Ref.~\cite{ahadi_strain-induced_2025}. Higher-order perturbation theory leads to additional terms with the same matrix structure and larger energy denominators, i.e., their contribution is further suppressed. Equation~\eqref{eq:delta_V1} corresponds to a renormalization of the mass parameter $M_0$ already present in the original model in Eq.~\eqref{eq:basic_CdAs}. Notice that our result for $\delta \mathcal{V}_1$ in Eq.~\eqref{eq:delta_V1} is not obtained by the simple replacement $A_{z}k_{z}\sigma_{x}\tau_{x}\rightarrow A_{z}(n\pi/L)\sigma_{x}\tau_{x}$~\cite{LiuDai_PRB_2025}, which incorrectly leads to a purely off-diagonal matrix structure in orbital and spin space. Such a term would even break inversion symmetry and thus not be consistent with the underlying symmetry of the system.

The replacement of $k_z \rightarrow n \pi/L$ is only allowed for terms that contain even powers of $k_z$. In that case, the first order in perturbation theory does not vanish. Considering a generic term of the form 
$\mathcal{V}_{2\ell}=A_{2\ell}\left(\hat{k}_{z}\right)^{2\ell}\sigma_{a}\tau_{b}$ with constant $A_{2\ell}$ and integer $\ell$, the leading first-order effect is indeed the replacement
\begin{equation}
    \mathcal{V}_{2}\rightarrow A_{2\ell}\left(\frac{n\pi}{L}\right)^{2\ell}\sigma_{a}\tau_{b}\,.
\end{equation}
This will be relevant when we consider models for strained samples in later sections, where a quadratic term $k_{z}^2\sigma_{y}\tau_{x}$ is allowed when both inversion and $C_{4}$ symmetries are broken. Given these considerations, in what follows, we do not fix the subband index and instead consider multiple subbands in our numerical calculations.
 
\subsection{Effects of out-of-plane  magnetic fields}
\label{subsec:out_of_plane_fields_sec_2}
In the presence of an orbital magnetic field $\mathbf{B} = B \hat{\mathbf{z}}$ along the $z$ direction, perpendicular to the film, we define the raising and lowering operators as~\cite{Goerbig-Les_Houches_Lecture_Notes_2009, RevModPhys.83.1193}
\begin{subequations}
\begin{align}
    a^\dagger &= \frac{l_B}{\sqrt{2}}\left( k_x - \nabla_y + \frac{y}{l_B^2}\right),\\
    a &= \frac{l_B}{\sqrt{2}}\left( k_x + \nabla_y + \frac{y}{l_B^2}\right).
\end{align}
\label{eqs:ladder_ops}
\end{subequations}
Here, $l_{B}=\sqrt{\frac{\hbar}{eB}}$ is the magnetic length with $e$ being the elementary charge and $\hbar$ the reduced Planck constant. In terms of $a$ and $a^\dagger$ and using $k_{x}=(k_{+}+k_{-})/2$ and $k_{y}=i(k_{-}-k_{+})/2$, we can make the following replacements,
\begin{subequations}
\begin{align}
    k_+ &\rightarrow \frac{\sqrt{2}}{l_B}a,\\
    k_- &\rightarrow \frac{\sqrt{2}}{l_B}a^\dagger,\\
   \frac{1}{2}\left(k_{+}k_{-}+k_{-}k_{+}\right) &\rightarrow \frac{2}{l_B^2}\left(a^\dagger a + \frac{1}{2}\right).
\end{align}
\label{eq:k_a}
\end{subequations}

The Hamiltonian 
for a given subband $n$ in an out-of-plane magnetic field then becomes
\begin{widetext}
\begin{align}
    \mathcal{H}^{\text{4/mmm}}_n(B \hat{\mathbf{z}}) &= \left( \begin{array}{cccc}
        M_n - \omega_{c}\left(a^\dagger a + \frac{1}{2} \right) & \frac{\sqrt{2}A}{l_B}a & 0 & 0\\
        \frac{\sqrt{2}A}{l_B}a^\dagger & - M_n + \omega_{c}\left(a^\dagger a + \frac{1}{2} \right) & 0 & 0\\
        0 & 0 & M_n - \omega_{c}\left(a^\dagger a + \frac{1}{2} \right) & - \frac{\sqrt{2}A}{l_B}a^\dagger\\
        0 & 0 & -\frac{\sqrt{2}A}{l_B}a & - M_n +  \omega_{c}\left(a^\dagger a + \frac{1}{2} \right)
    \end{array}\right),
\label{eq:UnStrainedModel}
\end{align}
\end{widetext}
where $\omega_{c} = 2M_{xy}/l_B^2$ is the cyclotron frequency. The zeroth (or lowest) Landau levels (LLLs) are exact eigenstates of Eq.~\eqref{eq:UnStrainedModel}, with wavefunctions
\begin{subequations}
    \begin{align}
    \psi_+(\nu=0) &= (0,0,\ket 0, 0)^T,\\
    \psi_-(\nu = 0) &= (0, \ket 0, 0, 0)^T. 
\end{align}
\label{eq:LL_zeroth}
\end{subequations}
 Here $\ket 0$ denotes the LLL, $\nu=0$, satisfying $a\ket 0=0$. The energies of $\psi_{\pm}(\nu = 0)$ are 
 \begin{equation}
     \epsilon_{\pm}(\nu=0)=\pm\left(M_{n}-\frac{\omega_{c}}{2}\right) \,.
 \end{equation}
 It is clear, therefore, that by changing the magnetic field, the LLs will cross at $\omega_{c}=2M_{n}$. In contrast, as we show now, the LLL will not cross if the bands are uninverted and thus trivial (see Eq.~\eqref{eq:lll_crossing_omega_c} below). 
 
 Having the exact form of these states without strain is useful to test whether a gap is trivially opened in zeroth-order perturbation theory by terms that are allowed in strained models with lower symmetry. In general, we will perform a systematic perturbative analysis, which consists of the following steps: (i) solve the unperturbed problem for the 4/mmm point symmetry up to quadratic order in $\mathbf{k}$; (ii) for reduced symmetries, use perturbation theory to integrate out all $m\neq n$ subbands and find an effective Hamiltonian for the $n$ subband (usually we focus on $n=2$); (iii) project all new terms using Eq.~\eqref{eq:LL_zeroth}.
 
 The projection (iii) of the newly allowed terms is performed onto the manifold generated by the two LLL states $\psi_\pm(\nu = 0)$. Note that this manifold corresponds to the subspace spanned by the states $\left|P,3/2\right\rangle $ and
$\left|S,-1/2\right\rangle $. In the following, we introduce Pauli matrices $\gamma_{\mu}$ in this subspace and the unstrained Hamiltonian projected onto this subspace reads
\begin{equation}
\tilde{\mathcal{H}}^{\text{4/mmm}}_n(B) =-\left[M_{n}-\omega_{c}\left(a^{\dagger}a+\frac{1}{2}\right)\right]\gamma_{z}.
\end{equation}

In later sections, when considering strained models, we project the effective Hamiltonians onto the subspace spanned by the LLL states in Eq.~\eqref{eq:LL_zeroth}, corresponding to a zeroth-order treatment in perturbation theory of the newly allowed terms. There, we show that different strain profiles lead to effective Hamiltonians of the form
\begin{equation}
\mathcal{H}_{\mathrm{LLL}}
=C_0\,\gamma_0 +\left(\frac{\omega_c}{2}-M_n - \Delta M_n\right)\gamma_z
  + C_x\,\gamma_x + C_y\,\gamma_y,
\label{eq:H_LLL} 
\end{equation}
where the parameters $C_{0},\Delta M_n, C_{x},C_{y}$ are functions of the magnetic field $B$ and of the band parameters of the newly allowed terms in the presence of strain. The effective Hamiltonian $\mathcal{H}_{LLL}$ has a dispersion of the form
\begin{equation}
\epsilon_{\text{LLL}}^{\pm}=C_{0}\pm\sqrt{\left[\frac{\omega_{c}}{2}-\left(M_{n}+\Delta M_{n}\right)\right]^{2}+C_{x}^{2}+C_{y}^{2}}.
\end{equation}
The term $C_{0}$ only leads to an unimportant vertical shift of the energies and will be neglected. The term $\Delta M_{n}$ leads to a renormalization of the mass $M_{n}$. The crossing of the LLL will only appear if $C_{x}=C_{y}=0$ and is then located at
\begin{equation}
\label{eq:lll_crossing_omega_c}
\frac{\omega_{c}}{2}=M_{n}+\Delta M_{n}\,,
\end{equation}
corresponding to the critical magnetic field strength 
\begin{equation}
B_{c}=B_{c}^{\left(0\right)}+\frac{\hbar}{e}\frac{\Delta M_n}{M_{xy}}, \label{eq:LL_moving}
\end{equation}
where $B_{c}^{\left(0\right)}$ is the critical field without strain. The crossing will only be present as long as $B_c>0$. In contrast, nonzero $C_x$ or $C_y$ leads to the opening of a gap between the LLL. As this is a perturbative argument, even if $C_x = C_y = 0$, there is no guarantee that the crossing will survive; to address this, we thus complement our analysis with exact diagonalization, where we keep a sufficiently large number of subbands. 
 
\subsection{Trivializing bands by tensile biaxial strain}
\label{subsec:trivializing_bands}
 Now we discuss, using a perturbative argument, how tensile strain may be sufficient to change the sign of the subband mass $M_n$, leading to a topological transition. A topologically trivial subband will naturally lead to a gap in the LL spectrum, as discussed when the magnetic field is included. This provides a possible explanation for the results of Ref.~\cite{ahadi_strain-induced_2025} that does not require any symmetry breaking.

We assume that the application of in-plane strain also changes the sample thickness $L = L_0 + \delta L$, $\delta L \ll L_0$ as determined by the Poisson ratio. The thickness-dependent mass of a subband $n$ is found from Eq.~\eqref{eq:Mass_n}, and in the strained sample, we can approximate
\begin{align}
     M_n &\approx M_{n,0} + 2M_z \left(\frac{n\pi}{L_0}\right)^2 \frac{\delta L}{L_0} \,.
     \label{eq:Mn_vs_deltaL}
\end{align}
Here, $M_{n,0}$ is the mass without strain. This equation can be solved for the case of $M_n = 0$, where the bands uninvert, corresponding to a topological phase transition into a phase with trivial bands. This leads to
\begin{align}
    \frac{\delta L}{L_{0}}=-\frac{1}{2}\frac{M_{n,0}}{M_{z}}\left(\frac{L_{0}}{n\pi}\right)^{2}.
    \label{eq:strain_deltaL}
\end{align}
Since $M_{z}$ is positive, this equation can only be fulfilled when $\delta L$ has the opposite sign compared to $M_{n,0}$.
Taking the reported {\it ab initio} bulk parameter values for the unstrained case, $M_0 = 28.2$~meV, $M_z = 207.2$ ~meV$\cdot$nm$^2$ from Ref.~\cite{villar_arribi_topological_2020}, we predict that the gap is inverted for $\delta L /L = -0.12 \equiv 12 \%$, or $\delta L=-2.32$~nm, as $M_{n=2} \approx 5.5$ meV for $L_0 = 19$ nm. Notice that within this approximation, we do not take into account possible modifications of $M_0$ and $M_z$ as a function of strain, which may lead to substantially lower strain fields required for gap inversion~\cite{villar_arribi_topological_2020}. 

An alternative view on the possibility of trivializing the bands via strain relies on {\it ab initio} calculations for compressive strain, reported in Ref.~\cite{villar_arribi_topological_2020}. There, the effect of strain on the Cd$_3$As$_2$ $k\cdot p$ model coefficients was investigated, and they report bulk values for the 
strained case,  $M_0 = 37.4$ meV, $M_z = 203.6$ meV$\cdot$nm$^2$ with $0.7$\% compressive strain ($\delta L>0$) along the $a-$axis. Notably, this corresponds to an $\approx 9$~meV increase of $M_0$ under this realistic strain size. Assuming that tensile strain ($\delta L<0$) changes $M_0$ by the same amount but with opposite sign, a topological phase transition to trivial bands is possible under the tensile strains applied in Ref.~\cite{ahadi_strain-induced_2025}. We note that we find within our $k\cdot p$ model calculation below that $M_0$ is indeed renormalized towards smaller values by terms that appear due to strain lowering the symmetries of the system (see Sec.~\ref{subsec:mmm}). Still, further work using {\it ab initio} calculations is needed to clarify this question. 

In any case, the inclusion of out-of-plane magnetic fields for a trivial band structure naturally leads to a gap in the LLL versus $B$ spectrum. Thus, tensile strain can potentially explain the experimental results of Ref.~\cite{ahadi_strain-induced_2025} by a pure band parameter renormalization effect without requiring any symmetry reduction induced by strain. 
 
\section{Biaxial Strain effects~\label{sec:biaxial_strain}}
In this and the following two sections, we derive $k\cdot p$ models for strained Cd$_3$As$_2$ thin films and analyze their behavior in magnetic fields, focusing on the fate of the crossing between the LLLs. Here, we consider in-plane biaxial strains that preserve $C_4$ symmetry, as realized experimentally in Ref.~\cite{pardue_controlling_2021, ahadi_strain-induced_2025}.  Such strain fields reduce the point group symmetry, as an inversion-breaking structure is energetically close to the inversion-preserving case. As shown by DFT simulations, in the presence of strain, when grown, the structure with lower symmetry is energetically favored.  This symmetry lowering is relevant to the bands at the $\Gamma$ point, from 4/mmm to 4mm by breaking inversion and the twofold rotations $C_{2x}$ and $C_{2y}$ (see Table~\ref{tab:symm_operat}). 

\begin{figure*}
    \centering
    \includegraphics[width=1\textwidth]{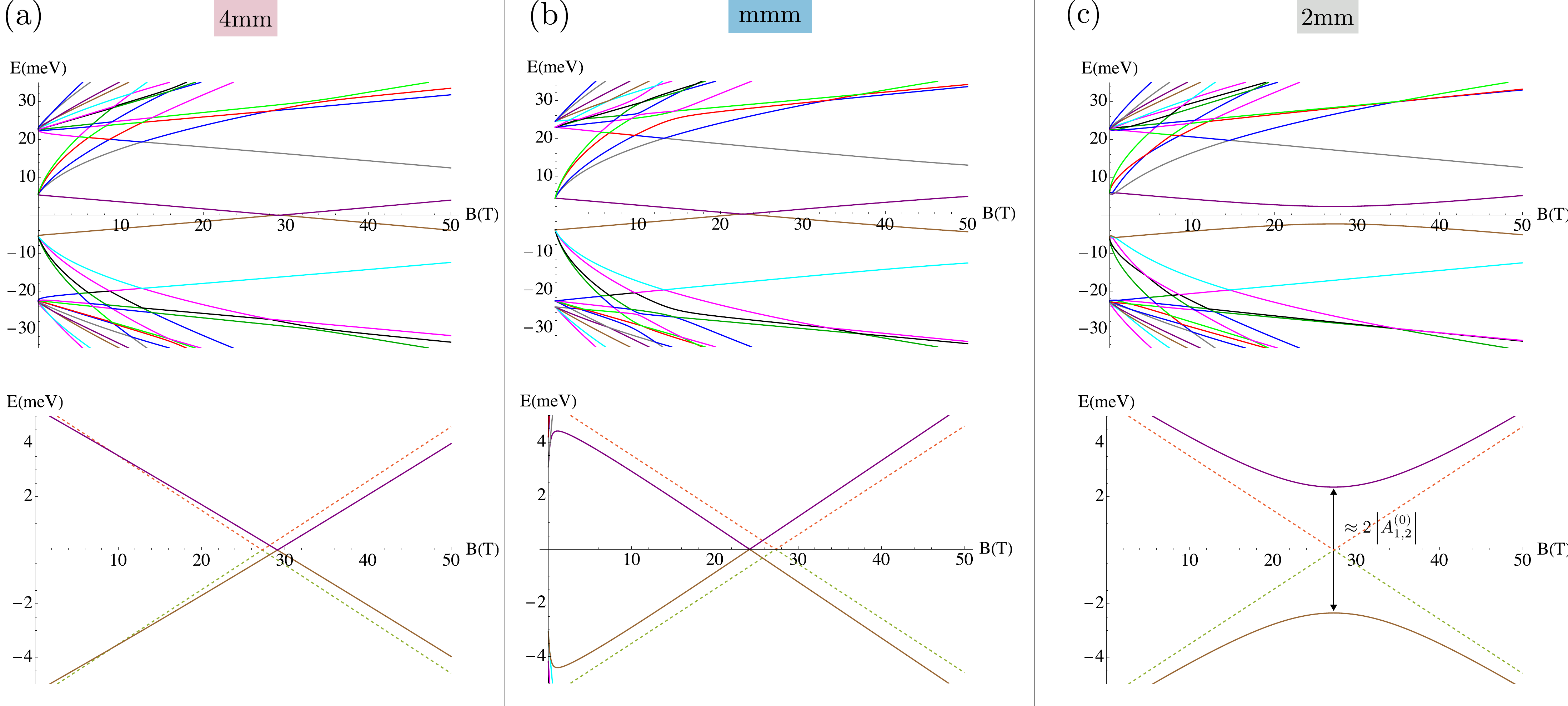}
    \caption{Effects of strain in a thin film with thickness $L=19\,\mathrm{nm}$. The $k \cdot p$ model parameters are $A=111.6\,\mathrm{meV\,nm}$, $M_{0}=28.2\,\mathrm{meV}$, $M_{z}=207.2\,\mathrm{meV\,nm^{2}}$, and $M_{xy}=133.2\,\mathrm{meV\,nm^{2}}$.  In all cases, we explicitly include three subbands and three LLs in the diagonalization, and we have explicitly checked that our conclusions are unchanged when using more. The top panels show all bands in the energy window from $-35$ to $35\,\mathrm{meV}$, while the bottom panels show a zoom from $-5$ to $5\,\mathrm{meV}$.  Without strain, the levels cross at $B\approx27\,\mathrm{T}$ (dashed lines in lower panels). (a) Biaxial strain with 4mm point symmetry that preserves $C_{4}$ rotations and breaks inversion, showing that the LLL crossing remains. We use $D_{s}=D_{p}=A/5=22.3\,\mathrm{meV\,nm}$ and $B_{1}=B_{2}=A_{2}=M_{xy}/5=26.6\,\mathrm{meV\,nm^{2}}$. (b) Spectra for mmm point symmetry, where $C_{4}$ is broken while inversion is preserved. The crossing remains, in sharp contrast with the naive replacement $k_{z}\to 2\pi/L$, highlighting the importance of including multiple subbands (or correctly incorporating them perturbatively) in the calculation. We choose $A_{z}=\delta A=A/5=22.3\,\mathrm{meV\,nm}$. We numerically verified that the result is even in $A_{z}$ (see Fig.~\ref{fig:field_shift}), consistent with perturbation theory. (c) Result for 2mm point symmetry, where both inversion and $C_{4}$ rotations are broken. Here, a momentum-independent off-diagonal term $\propto A_{1,2}^{(0)} \sigma_x \tau_y$ is allowed that opens a gap between the LLLs. We take this coupling to be $A_{1,2}^{\left(0\right)}=M_{0}/12=2.35\mathrm{meV}$. The perturbative calculation in the main text predicts a gap equal to $2A_{1,2}^{\left(0\right)}$ for small $A_{1,2}^{\left(0\right)}$ compared to the LL and subband gaps, which is numerically confirmed here.}
    \label{fig:LLsStrained}
\end{figure*}

\subsection{Thin film model with 4mm symmetry}
At lower 4mm symmetry, the Hamiltonian contains additional terms compared to the 4/mmm-symmetric model
\begin{equation}
    \mathcal{H}^{\text{4mm}} = \mathcal{H}^{\text{4/mmm}} + \Delta \mathcal{H}_{\text{4mm}}
\end{equation}
with 
\begin{equation}
\Delta\mathcal{H}_{4mm}=\Delta\mathcal{H}_{4mm}^{\left(1\right)}+\Delta\mathcal{H}_{4mm}^{\left(2\right)}\,.
\end{equation}
Here, we categorize the new terms according to their power in $(k_x, k_y, k_z)$.
There are no corrections to zeroth order, $\Delta\mathcal{H}_{\text{4mm}}^{\left(0\right)}=0$. The corrections that are linear in momentum read
\begin{align}
\Delta\mathcal{H}_{4mm}^{\left(1\right)} & =\frac{1}{2}\left(D_{s}+D_{p}\right)\left(k_{y}\sigma_{x}\tau_{0}-k_{x}\sigma_{y}\tau_{z}\right)+\nonumber \\
 & +\frac{1}{2}\left(D_{s}-D_{p}\right)\left(k_{y}\sigma_{x}\tau_{z}-k_{x}\sigma_{y}\tau_{0}\right)\,,
\end{align}
and the corrections quadratic in $\mathbf{k}$ read
\begin{align}
\Delta\mathcal{H}_{4mm}^{\left(2\right)} & =A_{2}k_{z}\left(k_{x}\sigma_{z}\tau_{y}+k_{y}\sigma_{0}\tau_{x}\right)\nonumber \\
 & -\frac{1}{2}\left(B_{1}-B_{2}\right)\left(k_{-}^{2}+k_{+}^{2}\right)\sigma_{x}\tau_{y} \nonumber \\
 & +\frac{i}{2}\left(B_{1}+B_{2}\right)\left(k_{+}^{2}-k_{-}^{2}\right)\sigma_{y}\tau_{y},
\end{align}
where $k_{\pm}=k_x\pm ik_y$. We choose these definitions of parameters $B_{1,2}$ and $D_{s,p}$ for later convenience when introducing the ladder operators to determine the Landau levels, as each coefficient multiplies a single ladder operator, $a$ or $a^\dagger$ [see Eqs.~\eqref{eq:k_a}]. 

Let us also state these linear and quadratic terms in $\mathbf{k}$ in their explicit matrix form, since it will be convenient when we discuss the effects of a magnetic field. They read 
\begin{subequations}
\label{eqs:BiaxialStrainTerms}
\begin{align}
\Delta H_{\text{4mm}}^{\left(1\right)} & =\left(\begin{array}{cccc}
0 & 0 & iD_{s}k_{-} & 0\\
0 & 0 & 0 & -iD_{p}k_{+}\\
-iD_{s}k_{+} & 0 & 0 & 0\\
0 & iD_{p}k_{-} & 0 & 0
\end{array}\right),\\
\Delta H_{\text{4mm}}^{\left(2,a\right)} & =A_{2}k_{z}\left(\begin{array}{cccc}
0 & -ik_{+} & 0 & 0\\
ik_{-} & 0 & 0 & 0\\
0 & 0 & 0 & ik_{-}\\
0 & 0 & -ik_{+} & 0
\end{array}\right),\\
\Delta H_{\text{4mm}}^{\left(2,b\right)} & =B_{1}\left(\begin{array}{cccc}
0 & 0 & 0 & -k_{-}^{2}\\
0 & 0 & k_{-}^{2} & 0\\
0 & k_{+}^{2} & 0 & 0\\
-k_{+}^{2} & 0 & 0 & 0
\end{array}\right),\\
\Delta H_{\text{4mm}}^{\left(2,c\right)} & =B_{2}\left(\begin{array}{cccc}
0 & 0 & 0 & -k_{+}^{2}\\
0 & 0 & k_{+}^{2} & 0\\
0 & k_{-}^{2} & 0 & 0\\
-k_{-}^{2} & 0 & 0 & 0
\end{array}\right).
\end{align}
\end{subequations}

\subsection{Out-of-plane magnetic field}
In the presence of an orbital magnetic field, we make the substitutions using Eqs.~\eqref{eq:k_a}. Since only terms that contain odd powers in $k_z$ couple different subbands, we separate the terms into two sets, depending on whether they couple the subbands ($\Delta H_{4mm}^{\left(2,a\right)}$) or are diagonal in the subband index (the three others). As before, we assume that the $n=2$ subband is the relevant one in Cd$_3$As$_2$ as it appears closest to the Fermi energy~\cite{Smith_PRB_2024, ahadi_strain-induced_2025}. Up to second order in perturbation theory, as shown in  Appendix~\ref{sec:App_Subbands}, we find an additional effective Hamiltonian term for the $n=2$ subband arising from $\Delta H_{4mm}^{\left(2,a\right)}$ that takes the form
\begin{equation}
    \Delta H_{\text{4mm}}^{\left(2,a\right)}
\propto -\,\frac{A_2^{\,2}}{ M_z}\,\big(k_x^2 + k_y^2\big)\,\sigma_0 \tau_z.
\end{equation}
We note that this form holds for any subband $n$, with possibly different pre-factors. 
This term simply amounts to a renormalization of the mass $M_0$ and does not open a gap in the LLL versus B spectrum, unless the renormalization is so strong that the mass term changes sign and the bands become topologically trivial (see Sec.~\ref{subsec:trivializing_bands}). It does, however, change the range of thicknesses $L$ in which the field-induced semimetal phase can be found (see Eq.~\eqref{eq:lll_crossing_omega_c}).  In fact, we can consider a zeroth-order argument in perturbation theory to see if a gap is immediately opened. For that, we consider the states of Eq.~\eqref{eq:LL_zeroth} and compute the projections described in Eq.~\eqref{eq:H_LLL}. The projection of all terms in Eq.~\eqref{eqs:BiaxialStrainTerms} vanishes, strongly suggesting that the crossing of the LLLs remains. 

To consider all terms exactly, we numerically diagonalize the Hamiltonian for a thin film of thickness $L=19$ nm, comparing the unstrained case with the case with biaxial strain. For the unstrained model, we consider the parameters from the {\it ab initio} work of Ref.~\cite{villar_arribi_topological_2020}: $A= 111.6$ meV nm, $M_0=28.2$ meV, $M_z=207.2$ $\text{meV nm}^{2}$, and $M_{xy}=133.2$ $\text{meV nm}^{2}$. For the exact diagonalization calculation, we always take $n_{\text{max}}=3$ subbands into account for each magnetic field strength $B$ and truncate to the three lowest LLs ($\nu = 0, 1,2$). We have explicitly checked that our conclusions are unchanged when taking more subbands and LLs into account. Without strain, the crossing appears close to $B\approx27T$. For the new strain-dependent terms in the 4mm model, we consider the generic values $D_s=D_p=A/5$ and $B_1=B_2=A_{2}=M_{xy}/5$. In Fig.~\ref{fig:LLsStrained}\hyperlink{fig:LLsStrained}{(a)}, we show exact diagonalization results of the spectra of the strained model (lines), superimposed with the spectra of the unstrained model (dashed). Importantly, we confirm the perturbative analysis that the crossing of the LLL remains, but its position changes. For the parameters we consider, it now appears at a slightly larger magnetic field strength of $B_c \approx 29$~T. We conclude that lowering the symmetry by biaxial strain to point group 4mm does not allow for new terms in the effective model that can remove the crossing of the LLL, even when considering the mixing between different subbands.

\section{Inversion-symmetric C$_4$-breaking  strain~\label{sec:C4_breaking}}

We now consider strain fields that break $C_4$ symmetry but preserve inversion, i.e. systems that exhibit mmm point group symmetry. For concreteness, we consider $B_{1g}$-type strain, for which according to Table~\ref{tab:symm_operat} the two diagonal mirrors $M_\pm$ are broken in addition to $C_4$. We have verified that $B_{2g}$-strain, which breaks the mirrors $M_x, M_y$ instead, leads to the same conclusions. The main finding of this section is that breaking $C_{4}$ is not sufficient to gap out the LLL crossing. 

\subsection{Thin film model with mmm symmetry}
\label{subsec:mmm}

We begin by writing the $k \cdot p $ model in the presence of $B_{1g}$ strain. The point group mmm is generated by $C_{2z}$, $C_{2x}$ and inversion, which allow the following corrections
to the 4/mmm model
\begin{equation}
    \mathcal{H}^{\text{mmm}} = \mathcal{H}^{\text{4/mmm}} + \Delta \mathcal{H}^{\text{mmm}}_{B_{1g}}
\end{equation}
with 
\begin{align}
\Delta\mathcal{H}_{B_{1g}}^{\text{mmm}} & =\Delta\mathcal{H}_{B_{1g}}^{\left(1\right)}+\Delta\mathcal{H}_{B_{1g}}^{\left(2\right)}+\Delta\mathcal{H}_{B_{1g}}^{\left(3\right)}. \label{eq:general_C4_I_break}
\end{align}
Sorting the terms according to their power in $(k_x, k_y, k_z)$ again, we first observe that 
there are no new constant terms $\Delta\mathcal{H}_{B_{1g}}^{\left(0\right)}=0$.
This has important implications for the question of whether the LLL crossing is gapped out, as we discuss later in this section. At linear order, the in-plane dispersion becomes anisotropic, and an additional linear term in $k_{z}$, which couples different subbands, is also symmetry allowed,
\begin{equation}
\Delta\mathcal{H}_{B_{1g}}^{\left(1\right)}=\delta A\,\left(k_{x}\sigma_{z}\tau_{x}+k_{y}\sigma_{0}\tau_{y}\right)+A_{z}k_{z}\sigma_{x}\tau_{x}.
\end{equation}
We note that a term linear in $k_{z}$ of this form was also considered in Ref.~\cite{villar_arribi_topological_2020}. As discussed in previous sections, the linear terms in $k_{z}$ couple different subbands, which we treat both analytically within perturbation theory (see Appendix~\ref{sec:App_Subbands} for details) and also exactly via an exact numerical diagonalization that takes a sufficient number of subbands into account.

At quadratic order in $\textbf{k} = (k_x, k_y)$, the diagonal identity term and the mass term become anisotropic in the $k_x$-$k_y$ plane, leading to a correction of the form
\begin{equation}
\Delta\mathcal{H}_{B_{1g}}^{\left(2\right)}=\delta\epsilon_{B_{1g}}\left(\mathbf{k}\right)\sigma_{0}\tau_{0}+\delta M_{B_{1g}}\left(\mathbf{k}\right)\sigma_{0}\tau_{z},
\end{equation}
with
\begin{align}
\delta\epsilon_{B_{1g}}\left(\mathbf{k}\right) & =\delta\epsilon\left(k_{x}^{2}-k_{y}^{2}\right),\\
\delta M_{B_{1g}}\left(\mathbf{k}\right) & =\delta M_{xy}\left(k_{x}^{2}-k_{y}^{2}\right).
\end{align}
These terms arise because $C_{4}$ invariance is no longer present. Since the matrix structure of these terms was already present in the unstrained 4/mmm symmetric model, they can be neglected in a first approximation, as they will not affect the question of whether the LLL crossing remains. 

\subsection{Out-of-plane magnetic field}
The new terms that are linear in $(k_x, k_y, k_z)$, that is $\Delta\mathcal{H}_{B_{1g}}^{\left(1\right)}$, are written in matrix form as
\begin{subequations}
\label{eqs:C4BreakingStrainTerms}
\begin{align}
\Delta\mathcal{H}_{B_{1g}}^{(1,a)}&=\delta A\left(\begin{array}{cccc}
0 & k_{-} & 0 & 0\\
k_{+} & 0 & 0 & 0\\
0 & 0 & 0 & -k_{+}\\
0 & 0 & -k_{-} & 0
\end{array}\right),\\
    \Delta\mathcal{H}_{B_{1g}}^{(1,b)} &= A_{z}k_{z}\sigma_{x}\tau_{x}=  A_{z}k_{z}\left(\begin{array}{cccc}
0 & 0 & 0 & 1\\
0 & 0 & 1 & 0\\
0 & 1 & 0 & 0\\
1 & 0 & 0 & 0
\end{array} \right).
\end{align}
\end{subequations}

We start by analyzing the effects of these terms perturbatively, using the states of Eq.~\eqref{eq:LL_zeroth} and projecting them using  Eq.~\eqref{eq:H_LLL}. The term $\Delta\mathcal{H}_{B_{1g}}^{(1,a)}$ is trivial to project, leading to a vanishing result. The term  $\Delta\mathcal{H}_{B_{1g}}^{(1,b)}$ would open a gap if $k_z$ is simply replaced by $n\pi/L$. However, as mentioned previously in Sec.~\ref{subsec:out_of_plane_fields_sec_2}, this replacement is incorrect and explicitly breaks inversion. When the coupling between different subbands is incorporated correctly (as discussed in detail in App.~\ref{sec:App_Subbands}), we find within perturbation theory that $\Delta\mathcal{H}_{B_{1g}}^{(1,b)}$  renormalizes the mass term $M_0$, preserves inversion, and only shifts the position of the LLL crossing but does not open a gap. The situation for point group symmetry mmm is therefore analogous to the $C_{4}$-preserving case of 4mm. 
This result is in contrast with previous studies~\cite{LiuDai_PRB_2025}, which concluded that the terms that arise from breaking $C_4$ (while preserving inversion) would lead to the opening of a gap in the LLL vs. B spectrum. We find, however, by performing the projection and using Eq.~\eqref{eq:LL_moving} that
\begin{equation}
    B_{c}\approx B_{c}^{\left(0\right)}-\alpha\frac{\hbar}{e}\frac{A_{z}^{2}}{M_{xy}M_{z}}, \label{eq:Bc_Az}
\end{equation}
with $\alpha$ a dimensionless number of order one. This shows, indeed, a predicted shift of the magnetic field for this crossing.

\begin{figure}
    \centering \includegraphics[width=0.5\textwidth]{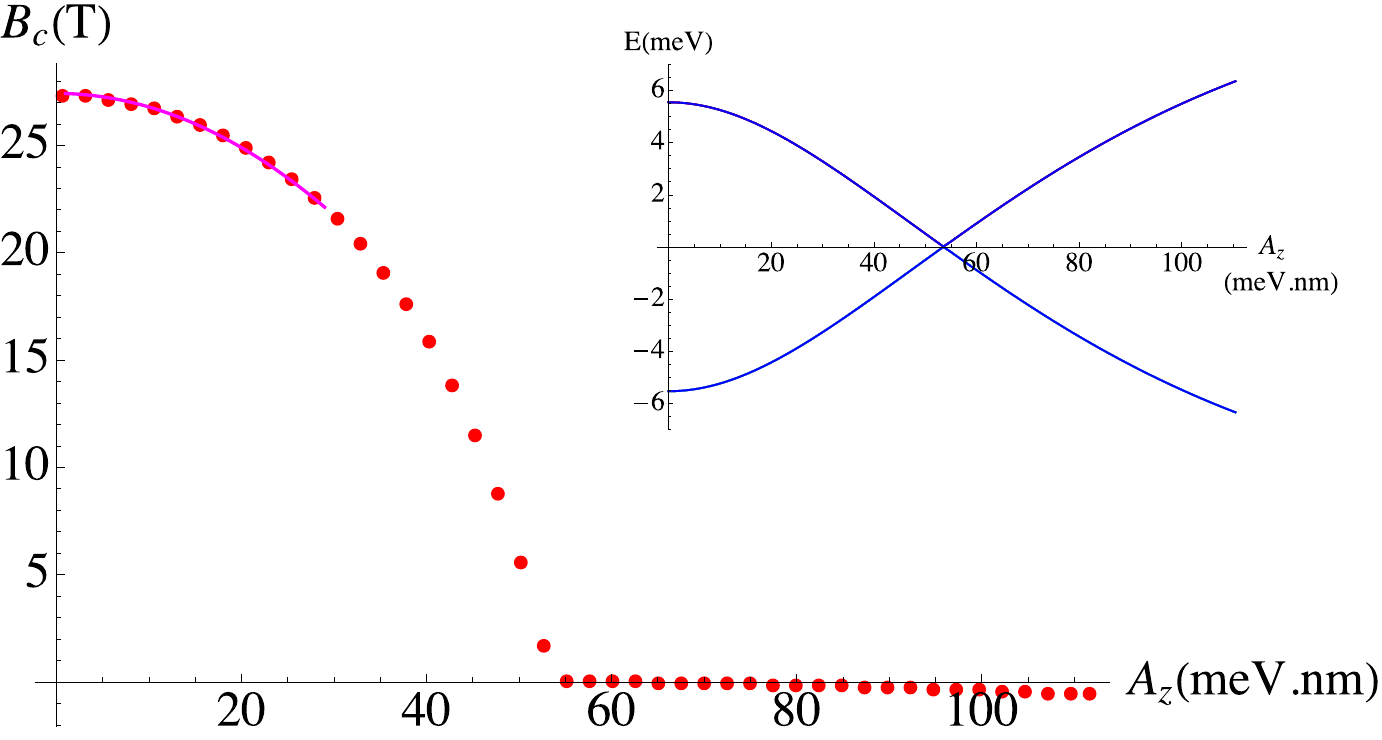}
    \caption{Dependence of the critical field $B_c$, where the crossing of LLLs occurs, as a function of $A_z$. The result (dots) is obtained from exact diagonalization considering three subbands. For clarity, we kept all the parameters identical to the unstrained case, and set $\delta A=0$ to pinpoint the effects of the $A_z$ term. The curve for small $A_z$ corresponds to a quadratic fit motivated by the perturbative result in Eq.~\eqref{eq:Bc_Az}. 
    The critical field vanishes for $A_z = 53$~meV nm, and for even larger values, no crossing will occur as the respective subbands have uninverted and become trivial. Inset: The corresponding low-energy bands as a function of $A_{z}$, showing that the gap closes when $B_c$ reaches zero, pointing to a topological phase transition that trivializes the relevant bands. 
    }
    \label{fig:field_shift}
\end{figure}

To complement the analytical perturbative calculation, we perform a numerical exact diagonalization, choosing $A_{z}=\delta A=A/5 = 22.3$~meV nm, and keeping the same values of the other parameters as in the unstrained case. We keep the $n=1,2,3$ subbands and $\nu = 0, 1, 2$ LLs, and have checked that the results are unchanged when keeping more subbands or LLs. The results are shown in Fig.~\ref{fig:LLsStrained}\hyperlink{fig:LLsStrained}{(b)}. In Fig.~\ref{fig:field_shift}, we show the critical field dependence $B_c$ (i.e. the location of the LLL crossing) versus $A_z$ for fixed $\delta A$. We perform a fit with the form of Eq.~\eqref{eq:Bc_Az} and find a perfectly parabolic behavior. The fit coefficient takes the value $\alpha\approx 0.11$, confirming the perturbative approach. Importantly, we observe that the critical field value $B_c$ decreases with increasing $A_z$. Since $A_z = 0$ in the unstrained sample and $|A_z|$ thus increases (at least to lowest order) with strain, we find that $B_c$ decreases with strain. This opens the possibility of reducing $B_c$ to zero, which corresponds to the removal of the crossing by sufficiently large strain, as the $n=2$ case then becomes trivial. This corresponds to the left, symmetry-preserving pathway in the diagram in Fig.~\ref{fig:CdAsStrain}\hyperlink{fig:CdAsStrain}{(c)}. 

\section{Breaking both C$_4$ and inversion symmetry \label{sec:break_C4_inversion}} 
In this section, we consider the effects of breaking both $C_{4}$ and inversion symmetries. The symmetry of the material is then captured by the point group $\text{2mm}$, which is generated by $C_{2z}$ rotation, and the mirrors $M_{x}$ and $M_{y}$. The symmetries of the system also include time reversal $T$ (see Table~\ref{tab:symm_operat}). Two distinct cases can physically lead to this case. The first case corresponds to biaxial strain with the addition of a $C_{4z}$ breaking symmetry field, i.e., small deviations from purely biaxial strain that slightly break $C_{4z}$ down to $C_{2z}$. Another case is uniaxial strain with broken inversion. This case is relevant, for instance, when placing the sample on a substrate that imposes uniaxial strain on one side of the sample only. 

 Several additional terms are allowed in $\text{2mm}$:
 \begin{equation}
     \mathcal{H}^{\text{2mm}} = \mathcal{H}^{\text{4/mmm}} + \Delta \mathcal{H}^{\text{2mm}} \,.
 \end{equation}We again classify them according to the powers of $(k_x, k_y, k_z)$ and find
\begin{equation}
\Delta\mathcal{H}^{\text{2mm}}=\Delta\mathcal{H}_{\text{2mm}}^{\left(0\right)}+\Delta\mathcal{H}_{\text{2mm}}^{\left(1\right)}+\Delta\mathcal{H}_{\text{2mm}}^{\left(2\right)}.
\end{equation}

At zeroth order in $\textbf{k}$, we find a constant term that will ultimately lead to a gap between the LLL, as we show later in this section. It reads
\begin{equation}
\Delta\mathcal{H}_{\text{2mm}}^{\left(0\right)}=A_{1,2}^{\left(0\right)}\sigma_{x}\tau_{y}. \label{eq:GapOpeningStrainTerm}
\end{equation}
We emphasize that this term appears only when both $C_{4}$ and inversion are broken. In previous studies~\cite{tarasenko_split_2015}, the sample interfaces led to an inversion asymmetry in HgTe/CdTe quantum wells, also leading to gapping out of the LLL versus B crossing.  
Here, we find that incorporating symmetry breaking directly into the point symmetry group naturally leads to the same effect, but now directly via an analysis of the symmetry-allowed terms in the effective $k \cdot p$ model. 

There are several new linear terms. They read
\begin{align}
\Delta\mathcal{H}_{\text{2mm}}^{\left(1\right)} & =k_{y}\left(A_{1,0}^{\left(1\right)}\sigma_{x}\tau_{0}+A_{1,3}^{\left(1\right)}\sigma_{x}\tau_{z}\right)+\nonumber \\
 &
+k_{x}\left(A_{2,0}^{\left(1\right)}\sigma_{y}\tau_{0}+A_{2,3}^{\left(1\right)}\sigma_{y}\tau_{z}\right) +
 \nonumber \\
 & +A_{z}k_{z}\sigma_{x}\tau_{x} 
+ \delta A\left(k_{x}\sigma_{z}\tau_{x}+k_{y}\sigma_{0}\tau_{y}\right)
\label{eq:linear_C2v}
\end{align}

These terms fall into three categories. The first are
the ones diagonal in orbital space $\tau$, with the couplings $A_{1,0}^{\left(1\right)},A_{1,3}^{\left(1\right)},A_{2,0}^{\left(1\right)}$
and $A_{2,3}^{\left(1\right)}$. All these terms break inversion and $C_{4}$ and, therefore, did not appear in previous sections. The second category is the term with coupling $A_z$, which is also allowed in mmm, as it is invariant under inversion. As discussed in the previous Sec.~\ref{sec:C4_breaking}, it couples different subbands since it is odd in $k_{z}$ and when correctly treated within perturbation theory (or fully numerically) simply renormalizes diagonal terms. Finally, we have the terms with coupling $\delta A$, which are also present in mmm, as they preserve inversion. 

Several new quadratic and cubic terms appear, and we will not list all of them here, as the main effects are already captured to linear order. We will, instead, focus on the most relevant ones. For instance, a quadratic term of the form $k_{z}^2\sigma_{x}\tau_{y}$ is allowed. A term of this form was analyzed in Section~\ref{sec:unstrained} and Appendix~\ref{sec:App_Subbands}. Perturbatively, it leads to a subband-dependent renormalization of the constant term in Eq.~\eqref{eq:GapOpeningStrainTerm}. 

\subsection{Out-of-plane magnetic field}
\label{subsec:2mm_out_of_plane_field}
We start by analyzing the effects of these terms perturbatively, projecting onto the manifold of LLL states in Eq.~\eqref{eq:LL_zeroth}, which yields an effective Hamiltonian of the form in Eq.~\eqref{eq:H_LLL}. The zeroth-order term $\Delta\mathcal{H}_{\text{2mm}}^{\left(0\right)}$ already opens a gap. By comparing with Eq.~\eqref{eq:H_LLL}, we find $C_{x}=-A_{1,2}^{\left(0\right)}$. This is sufficient to open a gap, and in fact, the gap will be largely dominated by this term. The terms that are linear in $(k_x, k_y, k_z)$ lead to a vanishing contribution in this order in perturbation theory, with the exception of $A_z$, which renormalizes the mass parameter $M_0$ (see Eq.~\eqref{eq:delta_V1}).

In Fig.~\ref{fig:LLsStrained}\hyperlink{fig:LLsStrained}{(c)}, we confirm this perturbative analysis using exact diagonalization and show that a finite value of $A_{1,2}^{\left(0\right)}$ leads to a gap that is proportional to $A_{1,2}^{\left(0\right)}$. We set $A_{1,2}^{\left(0\right)}=M_{0}/12$ and verify that the gap is well approximated by $2 |A_{1,2}^{\left(0\right)}|$, in agreement with the perturbative calculation.

\section{In-plane magnetic Fields}
\label{sec:inplane_fields}

\begin{figure}
    \centering
    \includegraphics[width=0.45\linewidth]{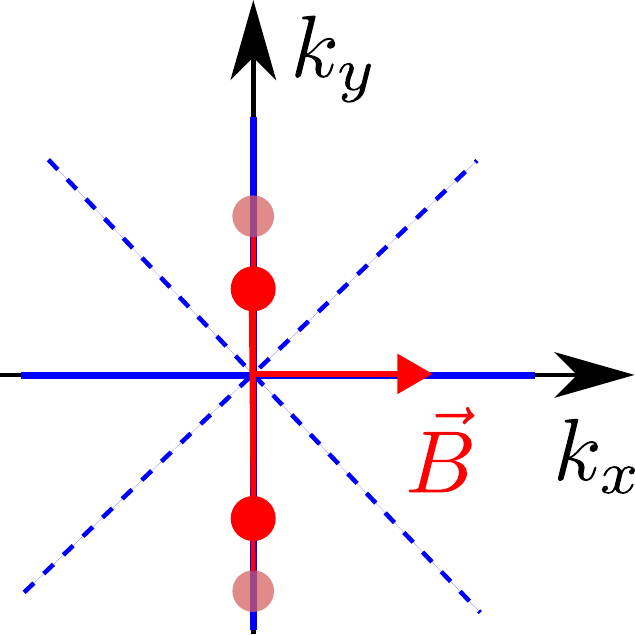}
    \caption{Emergence of the Dirac points at large in-plane magnetic fields and the effects of strain. The red arrow indicates the in-plane field direction, as determined by the polar angle $\phi$ (here $\phi = 0$). Dirac points emerge along $\theta_c = \phi \pm \frac{\pi}{2}$ whenever $\mathbf{B}$ points normal to a mirror plane. Strain shifts the location of the nodes along the mirror line, as schematically shown by red points (unstrained) and light red points (strained). Blue dashed lines indicate the diagonal mirrors, present only for 4/mmm and 4mm symmetry. The solid blue lines represent the horizontal and vertical mirrors present for all the symmetry groups considered in this work.
    }
    \label{fig:BandTouchingShift}
\end{figure}

In this section, we discuss the effects of strong in-plane magnetic fields. As shown in Ref.~\cite{Smith_PRB_2024}, in the absence of strain, the system transitions from an insulating state to a 2D Dirac semimetal for sufficiently large in-plane magnetic fields. For fields perpendicular to a mirror plane, the two Dirac points were shown to appear at an angle $\theta_{c}=\phi\pm\pi/2$, where $\phi$ is the direction of the in-plane field. This is schematically shown in Figure~\ref{fig:BandTouchingShift}. 

We will consider the different types of strain profiles and, for that, it is convenient to first consider a generic Hamiltonian for a given subband $n$ and later choose specific strain configurations. A generic subband Hamiltonian, up to linear order in $k$, can be written as
\begin{align}
\mathcal{H}_{n} & =\begin{pmatrix}  
M_{n} & A_{+} & \tilde{h}_{s}^{*} & F_{1}\\
A_{-} & -M_{n} & F_{2} & \tilde{h}_{p}^{*}\\
\tilde{h}_{s} & F_{2}^{*} & M_{n} & -A_{-}\\
F_{1}^{*} & \tilde{h}_{p} & -A_{+} & -M_{n}
\end{pmatrix}.\label{eq:H_n_Bpar}
\end{align}
As we show later, keeping terms up to linear order in $k$ is sufficient to capture all the main effects.

We focus on the low-energy model for the $n=2$ subband, which is relevant for Cd$_3$As$_2$ close to a thickness $L = 19$~nm. We express the magnetic field coupling to the $s$ and $p$-orbital electrons as $h_s e^{i \phi}$ and $h_p e^{3 i \phi}$,  where $h_s$ and $h_p$ are real positive amplitudes~\cite{Smith_PRB_2024}. Note that in the absence of strain and up to quadratic order in $\mathbf{k}$, we find $\tilde{h}_s = h_s e^{i \phi} $ and $\tilde{h}_p = h_p e^{3 i \phi}$. Here, $\phi$ is the polar angle of the magnetic field in the $xy$ plane. Given the indirect coupling of the $p$ electrons with in-plane fields, as the $M_{z}$ quantum numbers of the $p$ electrons differ by three (recall that $M_{z}=\pm 3/2$), we expect $h_{p} \ll h_{s}$. We also choose $\epsilon_0\approx-M_{2}$, and assume $|\tilde{h}_{s}|\gg\left|M_{2}\right|$, given that at the critical thickness $M_2 = 0$. 
In general, away from the critical thickness, the conversion of $M_2$ into an effective magnetic field comes from the Zeeman term. For a value of $M_2$ in meV, we can find the effective field corresponding to a value of $M_2$ as $M_2=\mu_{B} g\Delta M B$, with $M$ the azimuthal component of the total angular momentum. In this case, the transition is between states with $M=\pm3/2$, that is, $\Delta M=3$, while the Landé g-factor is $g=4/3$.  This leads to $B=4.32 M_2$, with $M_2$ in meV and $B$ in Tesla. For instance, for $L=19$ nm, $M_2=5.54$~meV and $B\approx 24$~T.

As in previous sections, we treat the off-diagonal, odd-power, $k_{z}$-dependent terms, which couple different subbands, perturbatively to obtain an effective model for a given subband. 

Solving for the $s-$orbital wavefunctions in terms of the $p-$orbital wavefunctions, we find the following effective Hamiltonian for the low-energy $p-$orbitals in the high-field limit, 
\begin{equation}
\mathcal{H}_{p}=\mathcal{H}_{pp}-\mathcal{H}_{ps}\,\mathcal{H}_{ss}^{-1}\,\mathcal{H}_{sp},
\end{equation}
where each term corresponds to a specific block of Eq.~\eqref{eq:H_n_Bpar}.
Specifically, we define
\begin{align}
\mathcal{H}_{pp} & =\begin{pmatrix}-M_{n} & \tilde{h}_{p}^{*}\\
\tilde{h}_{p} & -M_{n}
\end{pmatrix},\\
\mathcal{H}_{ss} & =\begin{pmatrix}M_{n} & \tilde{h}_{s}^{*}\\
\tilde{h}_{s} & M_{n}
\end{pmatrix},\\
\mathcal{H}_{sp} & =\begin{pmatrix}A_{+} & F_{1}\\
F_{2}^{*} & -A_{-}
\end{pmatrix},
\end{align}
with $H_{ps}=H_{sp}^{\dagger}$, where we used that $A_- = A_+^*$ since $\mathcal{H}_n$ is hermitian. Thus, within
the $p$-subspace and under the assumption of large fields, we find that the Hamiltonian for a given
subband reads
\begin{equation}
\mathcal{H}_{p}=-M_{n}+\left(\begin{array}{cc}
X_{1} & Y\\
Y^{*} & X_{2}
\end{array}\right),
\end{equation}
where
\begin{align}
X_{1} & =-\left(\frac{F_{2}A_{+}}{\tilde{h}_{s}^{*}}+\frac{F_{2}^{*}A_{-}}{\tilde{h}_{s}}\right),\\
X_{2} & =\frac{F_{1}A_{+}}{\tilde{h}_{s}^{*}}+\frac{F_{1}^{*}A_{-}}{\tilde{h}_{s}},\\
Y & =\tilde{h}_{p}^{*}+\frac{A_{-}^{2}}{\tilde{h}_{s}}-\frac{F_{1}F_{2}}{\tilde{h}_{s}^{*}}.
\end{align}
In general, a crossing of energy levels (i.e., a nodal point) requires that $Y=0$ and $X_{1}=X_{2}$. Since these are complex numbers, this is a strong requirement. When $X_{1}=X_{2}$ is fulfilled, as required by symmetry up to linear order in all cases considered here, the remaining condition for $Y=0$ reduces to
\begin{equation}
A_{-}^{2}=-\tilde{h}_{p}^{*}\tilde{h}_{s}\left(1-\frac{F_{1}F_{2}}{\tilde{h}_{p}^{*}\tilde{h}_{s}^{*}}\right).\label{eq:A_cond_fields}
\end{equation}

Without strain and keeping the model up to quadratic order, we find $X_{1}=X_{2}=F_{1}=F_{2}=0$,
$\tilde{h}_{p}=h_{p}e^{3i\phi}$, $\tilde{h}_{s}=h_{s}e^{i\phi}$ and $A_{\pm}=Ak_{\pm}$.
We parametrize the in-plane momentum as $k_{\pm}=ke^{\pm i \theta}$, with $k>0$, such that Eq.~\eqref{eq:A_cond_fields} becomes
\begin{equation}
A^{2}k^{2}=-e^{-2i(\phi-\theta)}h_{p}h_{s}.
\end{equation}
This implies that real solutions for $k$ exist only if $e^{- 2 i (\phi - \theta)} = -1$ or $\theta-\phi=n\pi/2$ with $n$ being odd. Going to higher orders in the $k\cdot p$ model, it was shown in Ref.~\cite{Smith_PRB_2024} that a more stringent condition is found: band touchings can only occur if the magnetic field points normal to a mirror plane. We will now show that the same result also holds with strain. As long as a mirror is not broken by the applied strain, band touchings emerge in the limit of large in-plane fields along the intersection of the mirror plane with the $xy$ plane.

To analyze the strained cases, where more terms in the $k \cdot p $ models are allowed, we keep the novel terms up to linear order in $\mathbf{k}$ to determine whether band touchings emerge along specific directions. 

\subsection{Dirac nodes for 4mm symmetry}
Let us start with the 4mm group, relevant to the biaxial strain case. Up to linear order in $\mathbf{k}$, one finds
\begin{align}
A_{\pm} & =Ak_{\pm},\\
F_{1,2} & =0,\\
\tilde{h}_{s} & =h_{s}e^{i\phi}-iD_{s}k_{+},\\
\tilde{h}_{p} & =h_{p}e^{3i\phi}+iD_{p}k_{-}.
\end{align}
From Eq.~(\ref{eq:A_cond_fields}) we conclude that
\begin{equation}
A^{2}k_{-}^{2}=-\left(h_{p}e^{-3i\phi}-iD_{p}k_{+}\right)\left(h_{s}e^{i\phi}-iD_{s}k_{+}\right),
\end{equation}
implying that
\begin{align}
A^{2}k^{2} & =-e^{2i\left(\theta-\phi\right)}\left(h_{p}-ie^{i\left(3\phi+\theta\right)}D_{p}k\right)\times\nonumber \\
 & \times\left(h_{s}-ie^{-i\left(\phi-\theta\right)}D_{s}k\right). \label{eq:k_4mm}
\end{align}
Considering the angles in the interval $0$ to $\pi$, this equation has solutions for real $k$ for the pairs 
\begin{equation}   \left(\theta,\phi\right)\in \left\{\left(0,\frac{\pi}{2}\right),\left(\frac{\pi}{4},\frac{3\pi}{4}\right),\left(\frac{\pi}{2},0\right), \left(\frac{3\pi}{4},\frac{\pi}{4}\right)\right\}, \, \label{eq:pairs_solutions}
\end{equation}
the same pairs of angles that were solutions without strain. We here assume that $D_{p},D_{s}$ are small compared to $h_{p},h_{s}$, and will thus only slightly change the absolute value of $k$. While Eq.~\eqref{eq:k_4mm} has a family of solutions other than the ones listed in Eq.~\eqref{eq:pairs_solutions}, we expect that terms of higher order in $k$ restrict the solutions to the ones in  Eq.~\eqref{eq:pairs_solutions}, analogously to the case without strain~\cite{Smith_PRB_2024}.

\subsection{Dirac nodes for mmm and 2mm symmetry}
Now, we consider mmm symmetry, which contains only two vertical mirror planes. For our choice of $B_{1g}$ strain, these are $M_x, M_y$ (see Table~\ref{tab:symm_operat}). For $B_{2g}$ strain, these would be the diagonal mirrors instead. The terms up to linear order read 
\begin{align}
A_{\pm} & =Ak_{\pm}+\delta Ak_{\mp},\\
F_{1,2} & =0,\\
\tilde{h}_{s} & =h_{s}e^{i\phi},\\
\tilde{h}_{p} & =h_{p}e^{3i\phi}.
\end{align}
From Eq.~(\ref{eq:A_cond_fields}) we find that
\begin{equation}
k^{2}\left(A^{2}+2A\delta Ae^{-2i\theta}+\left(\delta A\right)^{2}e^{-4i\theta}\right)=-e^{-2i\left(\phi-\theta\right)}h_{p}h_{s}.
\end{equation}
Real solutions thus exist for 
\begin{equation} \left(\theta,\phi\right)\in \left\{\left(0,\frac{\pi}{2}\right),\left(\frac{\pi}{2},0\right)\right\} \,.
\label{eq:angles_sols_mmm}
\end{equation}
with
\begin{equation}
k^2=\frac{h_{p}h_{s}}{A^{2}\pm2A\delta A+\left(\delta A\right)^{2}}.
\end{equation}
The signs $\pm$ correspond to the two possible pairs of angles in Eq.~\eqref{eq:angles_sols_mmm}. Notice that the other pairs of angles listed in Eq.~\eqref{eq:pairs_solutions} are no longer solutions, in contrast to the cases of 4/mmm or 4mm symmetry, where four vertical mirrors exist. We conclude, therefore, that the two remaining mirror planes $M_x, M_y$ protect the emergent Dirac nodes, whenever the field is applied along the coordinate axes, $\phi = 0, \pi/2$, i.e., along the mirror normals. Therefore, nodal band crossings can survive even in the presence of uniaxial strain.

Finally, for 2mm point symmetry, all terms that exist for 4mm and for mmm are allowed as well. Therefore, all conditions that we derived in these two cases have to be fulfilled. Thus, for $B_{1g}$ strain, Dirac nodes can thus only appear along the $x$ or $y$ axis, provided the magnetic field is applied along $y$ or $x$ axis, respectively. For $B_{2g}$ strain, they can appear along the diagonals, provided that $\phi = \pi/4, 3 \pi/4$. Considering, for instance, the zeroth-order in $\mathbf{k}$ term from Eq.~\eqref{eq:GapOpeningStrainTerm}, we find $F_{1}^{*}=F_{2}=iA_{1,2}^{\left(0\right)}$. This means that $X_{1}=X_{2}$ still holds. As the products $\tilde{h}_{p}^{*}\tilde{h}_{s}$ and $F_{1}F_{2}$ are both real for the choices of mirror planes along $x$ and $y$, as previously ensured, once again a real solution of Eq.~(\ref{eq:A_cond_fields}) can be found.

\section{Conclusions~\label{sec:Conclusions}}
In this paper, we derive effective low-energy models for (001) thin films of Cd$_3$As$_2$ in the presence of different types of strain fields and investigate their behavior in magnetic fields. We show that it is essential to correctly account for the coupling between different subbands, which arises from the confinement along $z$ in the thin film geometry. This coupling is mediated by terms containing odd powers of $k_z$, which we treat analytically within perturbation theory and by using exact numerical diagonalization. 

In perpendicular fields, we identify two possible routes towards removing a crossing of the LLL by strain. The crossing is important since it serves as a signature of the nontrivial topology of the relevant subband close to the Fermi energy. We find that the crossing can either be removed via a strain-induced renormalization of band parameters that uninverts the topological band structure, i.e. trivializes the relevant subband, or by breaking both fourfold rotation $C_4$ and inversion symmetry. The two scenarios can be distinguished by tuning the magnitude of the strain: while the critical field strength at the crossing is progressively reduced to zero in the scenario of trivializing the bands via strain, the gap opens immediately in the symmetry-breaking scenario. Our work thus provides a theoretical explanation of the results in Ref.~\cite{ahadi_strain-induced_2025} and shows that the observed absence of the LLL crossing implies that the bands must be trivialized by strain, provided the experimentally applied biaxial strain is indeed $C_4$-preserving. In fact, the relatively small band gap for the relevant subband, $M_{n=2} \approx 5.5$ meV for $19$~nm thin films, makes it completely plausible that strain tuning of the band parameters could uninvert the relevant bands. 

For in-plane magnetic fields, we show that breaking mirror planes has important consequences for the emergent 2D Dirac semimetal phase observed at large field strengths. Specifically, the Dirac semimetal acquires a gap if the in-plane field is applied perpendicular to mirror planes that are broken by the applied strain, therefore giving a separate diagnostic to confirm what symmetries are preserved. For example, comparing low-energy transport for fields directed along the axes ($x,y$) and along the diagonals can be used to experimentally test the breaking of $C_4$ symmetry, which implies the removal of two mirror planes and thus the gapping out of the Dirac nodes in one of the two configurations. In contrast, in the presence of $C_4$ symmetry, these two field directions should yield identical results, allowing one to distinguish 4/mmm and 4mm point symmetry from mmm and 2mm. 

A promising future direction would be to perform {\it ab initio} calculations to determine the relative size of the stress-generated mass term and enable quantitative agreement between theory and the experimental results. Also, {\it ab initio} results could provide parameter values for the different newly allowed terms in the low-energy models that we derive and that are generated by strain, allowing one to identify those that are of most relevance to the low-energy behavior of Cd$_3$As$_2$ thin films. It would also be interesting to extend our study to other thin film geometries of Cd$_3$As$_2$. Generally, the  systematic analytical treatment of the effects of strain that we provide, which includes multiple subbands and Landau levels, may be useful for the description of other thin film materials. 

\section{Acknowledgments}
We acknowledge useful discussions with Thaís V. Trevisan. M.S., A.A.B., and I.M. were supported by the Center for the Advancement of Topological Semimetals (CATS), an Energy Frontier Research Center funded by the U.S. Department of Energy (DOE) Office of Science (SC), Office of Basic Energy Sciences (BES), through the Ames National Laboratory under contract DE-AC02-07CH11358. Research at Perimeter Institute is supported in part by the Government of Canada through the Department of Innovation, Science and Economic Development and by the Province of Ontario through the Ministry of Economic Development, Job Creation and Trade. P.P.O. gratefully acknowledges financial support by the "Transformationsprogramm Forschung und Wissenstransfer Saar" through the Center for Quantum Technologies (QuTe). V.L.Q. acknowledges financial support from the CNPq grant 311565/2023-9, the University of São Paulo startup grant number 22.1.09345.01.2, and the Fapesp Grant no. 2024/09202-0. 

\appendix

\section{Perturbation theory in the subband basis~\label{sec:App_Subbands}}
In this appendix, we provide details of the perturbative approach explained in the main text. We are going to treat the problem without magnetic fields and at $k_x=k_y=0$.

\subsection{Unperturbed Hamiltonian}
The mass terms of the unperturbed Hamiltonian in coordinate representation, where $\hat{k}_{z}=-i\partial_{z}$
read
\begin{equation}
\mathcal{H}_{0}=\left[M_{0}-M_{z}\left(-i\partial_{z}\right)^{2}\right]\tau_{z}\sigma_{0}.\label{eq:H0_appendix}
\end{equation}
The eigenstates of \eqref{eq:H0_appendix} can be written as $\left|n,\tau,\sigma\right\rangle $, where $n$ labels the subbands, with the proper eigenfunctions written in the main text in Eqs.~\eqref{eq:subband_Psi_abn} and~\eqref{eq:subbands_eigenf}. Defining $k_{n}=\pi n/L$, we find the energies
\begin{equation}
E_{0}\left(n,\tau\right)=\tau\left[M_{0}-M_{z}\left(\frac{\pi n}{L}\right)^{2}\right],
\end{equation}
each doubly degenerate in spin, because they do not depend on $\sigma$. For the mass term to vanish and the gap to close, we need the following condition to be satisfied for some value of $n$, 
\begin{equation}
M_{0}-M_{z}\left(\frac{\pi n}{L}\right)^{2}=0.
\end{equation}
In the absence of strain, this condition is satisfied for $n=2$ and $L=L^{*}$, which leads to
\begin{equation}
M_{z}=M_{0}\left(\frac{L^{*}}{2\pi}\right)^{2}.\label{eq:Mz_crit_cond}
\end{equation}
In this case, the projected low-energy manifold at $k_x=k_y=0$ will be four-fold degenerate, spanned by $\left|n=2,\tau,\sigma\right\rangle $, with $E_{0}\left(n=2,\tau\right)=0$. For $L > L^*$ the $n=2$ subband is topological. 

We are interested in adding strain to connect the subbands. In what follows, we consider the general case and thus do not assume that $L=L^{*}$, unless stated otherwise. If the bulk $k \cdot p$ model contains terms that contain odd powers of $k_z$, different subbands are coupled. This off-diagonal coupling between subbands makes it difficult to gain some analytical understanding of the strain effects, and, therefore, a question that we address in this appendix is how to calculate the perturbative corrections to the Hamiltonian arising from the subband couplings. 

For that, the first step is to compute the matrix elements between subbands. First, we do not consider the $\sigma$ and $\tau$ labels, and we will focus only on the $\left|n\right\rangle $ states, which is possible as the subspaces are disjoint. The terms that connect the subbands are the ones proportional to $k_{z}=-i\partial_{z}$ in the coordinate representation. A generic term can be written as
\begin{align}
\left\langle n\left|\left(-i\partial_{z}\right)^{\ell}\right|m\right\rangle  & =\int_{0}^{L}dz\varphi_{n}^{*}\left(z\right)\left(-i\partial_{z}\right)^{\ell}\varphi_{m}\left(z\right),\\
 & =\left(-i\right)^{\ell}\int_{0}^{L}dz\varphi_{n}^{*}\left(z\right)\left(\frac{\partial^{\ell}}{\partial z^{\ell}}\right)\varphi_{m}\left(z\right).
\end{align}
The most relevant terms are the ones with $\ell=1$ and $\ell=2$, since the $k \cdot p$ model corresponds to a Taylor series around $k_z = 0$. We define $f_{\ell}(n,m)=\left\langle n\left|\left(-iL\partial_{z}\right)^{\ell}\right|m\right\rangle$. Straightforward integration over $z$ yields, using the form of $\varphi(z)$ from the main text, the following non-vanishing results
\begin{align}
f_{1}\left(n,m\right) & =\frac{-i 4mn}{n^{2}-m^{2}},\,\,\text{for $\,m+n$ being odd},\\
f_{2}\left(n,m\right)& =\left(n\pi\right)^{2},\,\,\text{for $\,m=n$}.
\end{align}
Also notice that for any even $\ell$, 
\begin{equation}
f_{\ell}\left(n,m\right)=\left(n\pi\right)^{\ell},\,\,\,\text{for $m=n$}.
\end{equation}
This implies that terms with even powers of $k_z$ do not mix different subbands. 

\subsection{Perturbation theory}
We now consider the effects of several relevant terms in perturbation theory. We briefly recap how to write the perturbative corrections using the notation of Ref.~\cite{lindgren1974BW}. For convenience, we define the resolvent operator
\begin{equation}
R=Q\frac{1}{E_{0}-H_{0}}Q,
\end{equation}
where $P$ projects onto the desired ground-state manifold while $\ensuremath{Q=1-P}$ is the complement of $P$. Also, $\ensuremath{E_{0}}$ denotes the (possibly) degenerate energies of $H_{0}$ in the $P$ subspace without the perturbation $\mathcal{V}$~\cite{lindgren1974BW}. The corrections to $\mathcal{H}_0$ in the ground state manifold are given by
\begin{align}
\Delta \mathcal{H}^{\left(1\right)} & =P\mathcal{V}P,\label{eq:DeltaH_1}\\
\Delta \mathcal{H}^{\left(2\right)} & =P\mathcal{V}R\mathcal{V}P,\\
\Delta \mathcal{H}^{\left(3\right)} & =P\mathcal{V}R\mathcal{V}R\mathcal{V}P-P\mathcal{V}R^{2}\mathcal{V}P\mathcal{V}P.
\end{align}

For our specific problem of Cd$_3$As$_2$ thin films with thickness $L > L^*(n=2)$, we are interested in the corrections to the Hamiltonian for the subspace of the $n=2$ subband. Thus, 
\begin{align}
P & =\sum_{\sigma,\tau}\left|2,\tau,\sigma\right\rangle \left\langle 2,\tau,\sigma\right|,\\
 & =\mathcal{P}_{2}\otimes1_{\sigma}\otimes1_{\tau}.
\end{align}
We defined $\mathcal{P}_{2}=\left|2\right\rangle \left\langle 2\right|$
and separated it from the $\sigma$ and $\tau$ subspaces. Explicitly, one finds
\begin{align}
Q & =\sum_{m\ne2,\sigma,\tau}\left|m,\tau,\sigma\right\rangle \left\langle m,\tau,\sigma\right|,\\
 & =\mathcal{Q}_2\otimes1_{\sigma}\otimes1_{\tau}.
\end{align}
Here, we defined $\mathcal{Q}_n=\sum_{m\ne n}\left|m\right\rangle \left\langle m\right|$
as the projector onto all subbands with $m\ne n$. 

\subsection{Perturbations with $\hat{k}_{z}^{\ell}$, with $\ell$ even}

We will start with the simpler case of $\ell$ even. In this case, the first order correction, Eq.~(\ref{eq:DeltaH_1}), will be non-zero.
Assuming a perturbation $\mathcal{V}_{2}$ of the form
\begin{equation}
\mathcal{V}_{2}=A_{\ell}\left(\hat{k}_{z}\right)^{\ell}\sigma_{a}\tau_{b},\label{eq:V2_general}
\end{equation}
where $a,b\in\left\{ 0,x,y,z\right\}$ (0 denotes the identity), $\ell$ even, and $A_{_{\ell}}$ is $z-$independent. Computing the first-order correction to the Hamiltonian,
\begin{align}
\Delta \mathcal{H}^{\left(1\right)} & =A_{\ell}\left(P\hat{k}_{z}^{\ell}\sigma_{a}\tau_{b}P\right),\nonumber \\
 & =A_{\ell}\left(\mathcal{P}_{2}\hat{k}_{z}^{\ell}\mathcal{P}_{2}\right)\sigma_{a}\tau_{b}, \nonumber \\
 & =A_{\ell}f_{\ell}\left(2,2\right)\sigma_{a}\tau_{b},\nonumber \\
 & =A_{\ell}\left(\frac{2\pi}{L}\right)^{\ell}\sigma_{a}\tau_{b}.
\end{align}
This shows that for $\ell$ even, one can replace the operator
$\hat{k}_{z}=\left(-i\partial_{z}\right)$ by $k_{2}$ up to first
order in perturbation theory. 

\subsection{Perturbations with $\hat{k}_{z}^{\ell}$, with $\ell$ odd}
Motivated by the model with broken $C_4$ rotations, we consider, for concreteness, the term allowed in the case of $mmm$ symmetry,
\begin{equation}
\mathcal{V}_{1}=A_{z}k_{z}\tau_{x}\sigma_{x}=A_{z}\left(-i\partial_{z}\right)\sigma_{x}\tau_{x}.\label{eq:V1}
\end{equation}
The first-order correction vanishes
\begin{equation}
\Delta \mathcal{H}^{\left(1\right)}=P\mathcal{V}_{1}P=0,
\end{equation}
as the operator $\left(-i\partial_{z}\right)$ does not connect states
with the same $n$. This conclusion is, therefore, independent of
$\tau$ and $\sigma$. The second-order correction to the Hamiltonian of subband $n$ reads, 
\begin{align*}
& \Delta H_n^{\left(2\right)} =P\mathcal{V}R\mathcal{V}P,\\
 & =A_{z}^{2}\left[\mathcal{P}_{n}\left(-i\partial_{z}\right)\mathcal{Q}_n\right]\sigma_{x}\tau_{x}\frac{1}{E_{0}-H_{0}}\left[\mathcal{Q}_n\left(-i\partial_{z}\right)\mathcal{P}_{n}\right]\sigma_{x}\tau_{x}.
\end{align*}
Now, focusing on the case $n=2$, relevant to Cd$_3$As$_2$ with thickness close to $L^*(n=2)$ and using that the degenerate energy of the $P$ manifold that corresponds to the subband $n=2$, and applying
the projectors $\mathcal{P}_{2}$ and $\mathcal{Q}$,
\begin{align}
\Delta \mathcal{H}^{\left(2\right)} & =\frac{A_{z}^{2}}{L^{2}}\sum_{m\ne2}f_{1}\left(2,m\right)f_{1}\left(m,2\right)\sigma_{x}\tau_{x}\frac{1}{E_{0}-H_{0}}\sigma_{x}\tau_{x},\nonumber \\
 & =\frac{A_{z}^{2}}{L^{2}}\sum_{m\,\text{odd}}\tau_{x}\sigma_{x}\frac{\left|f_{1}\left(2,m\right)\right|^{2}}{\left\{ M_{z}\left[-\left(\frac{2\pi}{L}\right)^{2}+\left(\frac{m\pi}{L}\right)^{2}\right]\tau_{z}\sigma_{0}\right\} }\tau_{x}\sigma_{x},\nonumber \\
 & =\frac{A_{z}^{2}}{M_{z}}\left(\frac{1}{2\pi}\right)^{2}\sigma_{0}\tau_{z}\sum_{m\,\text{odd}}\frac{\left|f_{1}\left(2,m\right)\right|^{2}}{\left[1-\left(\frac{m}{2}\right)^{2}\right]}.
\end{align}

This term represents a renormalization of the mass $M_0$ in $\mathcal{H}_0$ due to the $C_{4}$-breaking perturbation. We find, therefore,
\begin{equation}
\Delta \mathcal{H}^{\left(2\right)}=\left(\frac{1}{2\pi}\right)^{2}\frac{A_{z}^{2}}{M_{z}}u_{1}\tau_{z}\sigma_{0},
\end{equation}
where we defined
\begin{equation}
u_{\ell}=\sum_{m\ne2}\frac{\left|f_{\ell}\left(2,m\right)\right|^{2}}{\left[1-\left(\frac{m}{2}\right)^{2}\right]}.
\end{equation}
In particular, for $\ell=1$, $u_{1}=-\pi^2$, leading to a final result 
\begin{equation}
    \Delta H^{\left(2\right)}=-\frac{A_{z}^{2}}{4 M_{z}}\sigma_{0}\tau_{z}\,.
\end{equation}
In the main text, there are additional terms in the Hamiltonian (such as the one with coupling constant $A$) in Eq.~\eqref{eq:basic_CdAs}, which will change the pre-factor of this result. The scaling with $A_{z}^{2}$, on the other hand, is robust, as we verified by exact diagonalization.

This calculation already indicates the general approach to treating other terms of the form~\eqref{eq:V2_general}. As long as they are proportional to $\left(\hat{k}_{z}\right)^{\ell}$, with $\ell$ odd, the first-order perturbation theory will be zero. The second-order correction will be
\begin{equation}
\Delta \mathcal{H}^{\left(2\right)}=-\frac{u_{\ell} A_{\ell}^{2}}{M_{z}L^{2\ell}}\left(\frac{L}{2\pi}\right)^{2}\sigma_{0}\left(\tau_{b}\tau_{z}\tau_{b}\right)\,.
\end{equation}
Notice that $\tau_{b}\tau_{z}\tau_{b}=\tau_{z}$ if $b=0,z$ and $\tau_{b}\tau_{z}\tau_{b}=-\tau_{z}$
otherwise. This result is, therefore, quite different from replacing
the operator $\hat{k}_{z}=\left(-i\partial_{z}\right)$ by $k_{2}$ in Eq.~\eqref{eq:V1}.

In conclusion, the replacement $\hat{k}_{z}\rightarrow k_{n}$ for a given subband $n$ works at first order for $\ell$ even but not for $\ell$ odd. For $\ell$ odd, the first-order perturbation theory vanishes, and the second order produces additional corrections with a generally different $\tau$ and $\sigma$ structure.

\bibliography{CdAsBib_strain.bib}
\end{document}